# Reflected light from giant planets in habitable zones: Tapping into the power of the Cross-Correlation Function

Martins, J. H. C.[1,2,3]; Santos, N. C.[2,3]; Figueira, P.[2,3]; Melo, C.[1]

[1] European Southern Observatory, Casilla 19001, Santiago, Chile
[2] Instituto de Astrofísica e Ciências do Espaço, Universidade do Porto, CAUP, Rua das Estrelas, 4150-762 Porto, Portugal
[3] Departamento de Física e Astronomia, Faculdade de Ciências, Universidade do Porto, Rua do Campo Alegre, 4169-007 Porto, Portugal

**Abstract**:

The direct detection of reflected light from exoplanets is an excellent probe for the characterization of their atmospheres. The greatest challenge for this task is the low planet-to-star flux ratio, which even in the most favourable case is of the order of $10^{-4}$ in the optical. This ratio decreases even more for planets in their host's habitable zone, typically lower than $10^{-7}$. To reach the signal-to-noise level required for such detections, we propose to unleash the power of the Cross Correlation Function in combination with the collecting power of next generation observing facilities. The technique we propose has already yielded positive results by detecting the reflected spectral signature of 51 Pegasi b (see Martins et al. 2015).

In this work, we attempted to infer the number of hours required for the detection of several planets in their host's habitable zone using the aforementioned technique from theoretical EELT observations. Our results show that for 5 of the selected planets it should be possible to directly recover their reflected spectral signature.

# 1 Introduction

20 years have passed since the discovery of 51 Pegasi b (Mayor and Queloz 1995), the first exoplanet confirmed around a solar-type star. This planet was the prototype of a new class of planets not predicted by existing models of planetary formation: hot-Jupiters, giant planets in extremely short period orbits. This groundbreaking discovery challenged the accepted cannons regarding planetary systems architecture and led to the reformulation of planetary formation theories as well as the introduction of the planetary migration concept (Lin, Bodenheimer, and Richardson 1996). Since then, close to two thousand planets have been confirmed in over 1229 systems (Schneider et al. 2011),some of them with a degree of complexity that rival our own Solar System (e.g. HD10180 with 7 planets, Lovis et al. 2011). Exoplanets have been found in all flavors, with masses smaller than the mass of the Earth to tens of Jupiter masses, orbital periods of a few hours to thousands of years (Mayor et al. 2011; Schneider et al. 2011).One of the most important conclusions from planet hunting programs is that planets are ubiquitous around solar-type stars (Howard et al. 2010).

The next step in exoplanet studies is the characterization of the atmosphere of the detected planets, for which several techniques were developed. Transmission spectroscopy (e.g. Charbonneau et al. 2002, Knutson et al. 2014), relies on the filtering of the host star spectrum by the planetary atmosphere during a transit. Occultation photometry and spectroscopy measures the wavelength dependency of the depth of the occultation of a transiting planet to infer the planetary thermal emission (e.g. Snellen et al. 2010) and reflected infrared signals (e.g. Rodler et al. 2013). As a planet shows us alternatively its day/night sides, the measurement of the flux variation of a planet allows to reconstruct the planetary phase variation along its orbit and reconstruct its reflected signal (e.g. Knutson et al. 2009). High-resolution spectroscopy (R~100000) has become a powerful tool for the study of exoplanetary atmospheres. In particular, the application of the Cross Correlation Function to high-resolution infrared observations already permitted the identification of some molecules in exoplanetary atmospheres (e.g. CO in the atmosphere of HD209458 – Snellen et al .2010). These kind of studies will be essential towards the identification of bio-markers in exoplanetary atmopheres (e.g. Snellen et al. 2013) towards the detection of extraterrestrial life.

The direct detection of the reflected spectra of exoplanets in the optical is one of the paths being taken by researchers. Although at a lower planet-to-star flux ratio, the direct detection of the planet at optical wavelengths consists mainly in reflected light as the thermal emission of exoplanets at optical wavelenghts is negligible. The reflected spectral signal from an exoplanetary atmosphere should mimic the stellar signal, permitting to infer the planetary albedo, a quantity highly dependent on the atmospheric composition of the planet. Albeit difficult, albedo measurements of exoplanetary atmospheres are now being used to constrain current models (e.g. Cowan & Agol 2011, Demory 2014) and the composition (e.g. the presence of clouds in HD 189733 b, Barstow et al. 2014) of exoplanetary atmospheres. Currently, the statistics of exoplanet albedos are still an open debate: initial findings suggested that in the optical the albedo of hot-Jupiters should be very low (e.g. Ag<0.08 for HD209458b - Rowe et al. 2008) but more recently higher albedos have been recovered (e.g. the case of Kepler-7b with an albedo that can be as high as 0.5 – Garcia-Muñoz and Isaak 2015).

Several attempts to use high-resolution spectroscopy towards the detection of the reflected spectroscopic signal of exoplanets in the optical have been made (e.g. for Tau Boo b - Charbonneau et al. 1999, Collier Cameron et al. 2002), albeit with inconclusive results. Although more attempts have been made at the detection of the reflected spectra signature of exoplanets (e.g. Leigh et al. 2003, Rodler, Kürster & Henning 2010), a conclusive detection has eluded researchers for many years. Nonetheless, all these attempts are of great scientific value because they allowed to establish upper limits on the planet-to-star flux ratio for the studied planets. More recently Martins et al. 2013 proposed a technique that makes use of the Cross Correlation Function of high resolution spectra to amplify the minute planetary signal above the stellar noise. The application of this method to high-resolution spectra of the prototypical hot-Jupiter 51 Pegasi b permitted to recover the reflected spectral signature of the host star reflected on the planetary atmosphere and infer that the planet may be an inflated hot-Jupiter with a high albedo (Martins et al. 2015).

The method used by (Martins et al. 2013; 2015) to detect the high-resolution spectroscopic signature of reflected light from an exoplanet opens new perspectives into this field. However, so far it has only been applied to one case: a giant planet in a short period orbit (51 Pegasi b). In this paper we explore the possibility to apply it to planets in the Habitable Zone of their host stars. In particular, we explore the possibility that future high-resolution spectrographs on giant telescopes (e.g. the E-ELT) will allow such detections.

# 2  Detecting reflected light from exoplanets

## 2.1  The Problem

The biggest issue in detecting the optical light of a star reflected on its orbiting planet is the extremely low planet-to-star flux ratio. Assuming a circular orbit, this flux ratio can be estimated from

$$\frac{F_{Planet}}{F_{Star}} = A_g g(\alpha) \left(\frac{R_{Planet}}{a}\right)^2 \qquad (1)$$

where $A_g$ is the geometric albedo of the planet, $g(\alpha)$ the phase function, $R_{Planet}$ the radius of the planet and $a$ the semi-major axis of the orbit (see Seager 2010 for details). This yields a maximum planet-to-star flux ratio in the optical at opposition ($g(\alpha)=1$) of the order of $10^{-4}$, even for the largest planets ever detected in the closest orbits. Table 1 shows the expected flux ratio for several prototypical planets at different orbital distances, as well as for the Earth and Jupiter for comparison.

| Planet | Period [days] | a [AU] | Albedo | Radius [$R_{Earth}$] | Flux ratio | SN for 3-sigma |
|---|---|---|---|---|---|---|
| | | | Model planets | | | |
| hot-Jupiter | 3 | | 0.3 | 13 | 5.3 x 10$^{-5}$ | **5.6 x 10$^4$** |
| hot-Neptune | 2 | | 0.3 | 5.5 | 1.6 x 10$^{-5}$ | **1.9 x 10$^5$** |
| | | | Real planets | | | |
| 51 Peg b | 4.23 | 0.052 | 0.5[1] | 20.9 | | |
| 55 Cnc e | 0.74 | 0.016 | *0.3* | 2.04 | 9.1 x 10$^{-6}$ | **3.3 x 10$^5$** |
| Solar System | | | | | | |
| Earth | 365.25 | 1 | 0.29 | 1 | 5.3 x 10$^{-10}$ | **5.7 x 10$^9$** |
| Jupiter | 4331.86 | 5.2 | 0.52 | 11 | 4.2 x 10$^{-9}$ | **7.1 x 10$^8$** |

Table 1: Expected flux ratio for prototypical and real exoplanets in different orbits. The expected values for the Earth and Jupiter as observed from outside the Solar System are also given for comparison. All flux ratio values were computed at opposition, i.e. *g(α)=1*. Unknown albedo values were set to 0.3 (in italic). *References for planet parameters*: *51 Peg b* - Martins et al. 2015 ; *55 Cnc e* - Gillon et al. 2012; *Earth* and *Jupiter* - de Pater & Lissauer 2010 .

From Table 1 it is clear that the required signal-to-noise ratio for the direct detection of the reflected signal of exoplanets, even in the best-case scenario, is extremely high, at the limit of current generation observing facilities. To surpass this problem we propose a technique described in detail in Martins et al. 2013 . This technique makes use of the Cross Correlation Function to enhance the minute planetary signal and has been used with success to detect the reflected signal of 51 Pegasi b from HARPS observations (see Martins et al. 2015) .

## 2.2 The technique

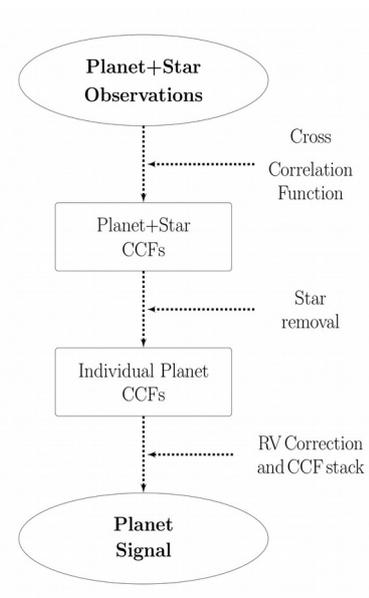

Figure 1: Schematic describing the detection method presented in (Martins et al. 2013)

The Cross Correlation Function (CCF for short) of a spectrum with a binary mask (e.g. Baranne et al. 1996) has been used successfully for many years on the detection of exoplanets via the radial velocity method. For a good approximation, it can be seen as stacking together all spectral lines from the mask identified in the spectrum to construct an average spectral line. Assuming that all lines have equal weight or that the weight is taken into account optimally (see Pepe et al. 2002), the gain in signal-to-noise from the spectrum to the CCF is proportional to the square root of spectral lines in the mask. For example, the HARPS binary mask for a G2 star has about 4100 spectral lines, which means that working with the CCF will correspond to an increase on S/N by a factor of ~65 when compared to working with the raw spectra.. This increase in S/N makes the CCF a powerful ally on the detection of the minute reflected optical signal of exoplanets.

Figure 1 shows how the CCF can be used to amplify the optical reflected signal from the planet to make it surface above the noise level. A detailed description of the process can be found in (Martins et al. 2013) and will not be replicated here.

This process was used successfully to recover the reflected optical light spectrum from the prototypical hot-Jupiter 51 Pegasi b (see Martins et al. 2015) at a significance of 3-sigma. Using this technique the authors were able to conclude that 51 Pegasi b is most likely an inflated hot-Jupiter with a high albedo (e.g., an albedo of 0.5 yields a radius of 1.9±0.3 R$_{Jup}$ for a signal amplitude of 6.0±0.4×10$^{-5}$). Since this is a direct observation on the planetary signal, they were also able to constrain the planet inclination (80$^{+10}_{-19}$ degrees) and real mass (M$_{51\ Peg\ b}$ = 0.46$^{+0.06}_{-0.01}$ M$_{Jup}$). These results are compatible with the work of Brogi et al. 2013 where they were able to infer a real mass of M$_{51\ Peg\ b}$ = 0.46±0.02 M$_{Jup}$ and an orbital inclination of *79.6 < I < 82.2* degrees from CRIRES observations.

The CCF of the recovered planetary signal can be seen in Figure 2.

---

[1]This value was estimated from the results of Martins et al. 2015 assuming a planet radius of 1.9 R$_{Jup}$.

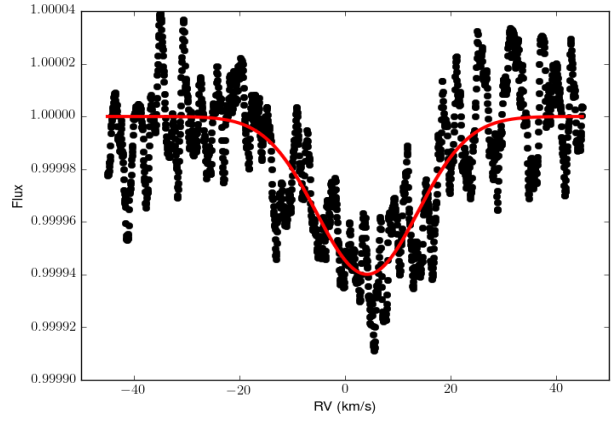

*Figure 2: CCF of the recovered planetary signal for 51 Pegasi b.*

# 3 Detection of possibly habitable zone planets

For this study, we created a sample of planets in the habitable zone of their host stars. To compute the HZ of each star, we adopted the same criteria of the Habitable Exoplanet Catalog (http://phl.upr.edu/hec), but without limiting the masses of the planet[2]. The computed HZ is a 1-D model based on the Kasting, Whitmire & Reynolds 1993 model. The inner $r_i$ and outer $r_o$ limits of the HZ are then given by:

$$\begin{aligned} r_i &= \left[ r_{is} - a_i (T_{eff} - T_s) - b_i (T_{eff} - T_s)^2 \right] \sqrt{L} \\ r_o &= \left[ r_{os} - a_o (T_{eff} - T_s) - b_o (T_{eff} - T_s)^2 \right] \sqrt{L} \end{aligned} \quad (2)$$

where $L$ is the stellar luminosity in solar units and $T_{eff}$ the effective temperature in K. The remaining factors are constants defined in the models of Selsis et al. 2007, Underwood, Jones & Sleep 2003: $T_s$ = 5700 K, $a_i$ = 2.7619e-5, $b_i$ = 3.8095e-9, $a_o$ = 1.3786e-4, $b_o$ = 1.4286e-9, $r_{is}$ = 0.72, and $r_{os}$ = 1.77.

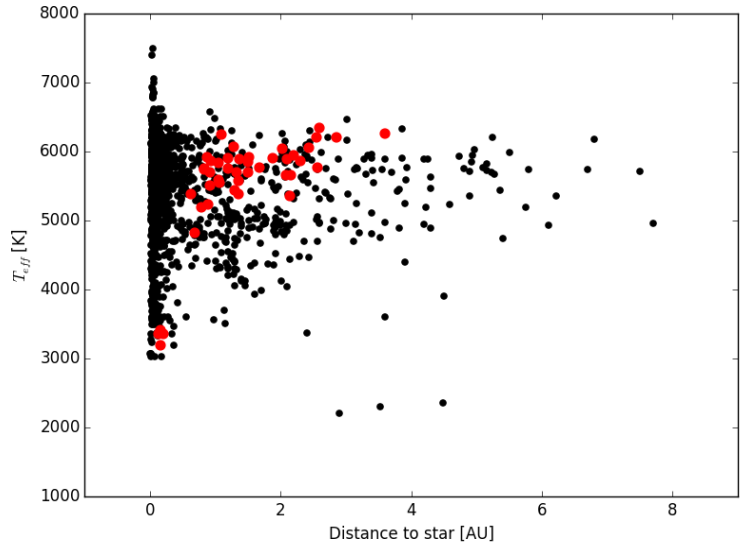

Figure 3: Planets in the Exoplanet.eu database (black dots) and the red dots are the planets in theirs host's HZ whose hosts are FJK stars on the main sequence.

---

[2]This will enable to encompass possible non-habitable giant planets with habitable satellites in the star.

The HZ was then computed for each planet host in the exoplanet.eu database (Schneider et al. 2011) from Equation 2.The final sample was restricted to planets in their host's HZ (i.e. with $r_i < a < r_o$) whose hosts are FJK stars on the main sequence (see Table 2). For those, we computed the maximum planet-to-star flux ratio as per Equation 1 assuming an albedo value of 0.5. For the planets whose radii have not been measured we used the empirical relation for planet radius, mass and incident flux[3] presented in Equation 3 (see Weiss et al. 2013 for details):

$$M_{Planet} < 150 M_{Earth} \quad \rightarrow \quad R_{Planet} = 1.78 \; (M_{Planet})^{0.53} \; (F_{incident})^{-0.03}$$

$$M_{Planet} \geq 150 M_{Earth} \quad \rightarrow \quad R_{Planet} = 2.45 \; (M_{Planet})^{-0.039} \; (F_{incident})^{0.094}$$

(3)

where the mass and radius of the planet are given in Earth units and the incident flux in c.g.s units. The incident flux is given by:

$$F = \sigma \; T_{eff}^4 \; \left(\frac{R_\star}{a}\right)^2 \sqrt{\frac{1}{1-e^2}} \qquad (4)$$

where all variables are in c.g.s units.

Figure 3 shows all planets in the exoplanet.eu database orbiting stars with effective temperatures between 2000K and 8000K (black dots). The red markers correspond to planets orbiting main sequence stars with $mag_V < 12$ that are within their host's habitable zones.

| Planet Name | Period [days] | a [AU] | Teff [K] | $r_i$ [AU] | $r_o$ [AU] | Mass [MJup] | Radius [RJup] | Flux Ratio |
|---|---|---|---|---|---|---|---|---|
| 16CygB b | 800 | 1.68 | 5766 | 0.70 | 1.72 | 1.7 | 0.6 | $1.4 \times 10^{-8}$ |
| 47Uma b | 1078 | 2.10 | 5892 | 0.92 | 2.25 | 2.5 | 0.6 | $8.3 \times 10^{-9}$ |
| 55Cnc f | 261 | 0.78 | 5196 | 0.56 | 1.40 | 0.1 | 0.8 | $1.1 \times 10^{-7}$ |
| Gl687 b | 38 | 0.16 | 3413 | 0.11 | 0.30 | 0.1 | 0.5 | $9.8 \times 10^{-7}$ |
| Gliese876 b | 61 | 0.21 | 3350 | 0.09 | 0.25 | 1.9 | 0.6 | $8.5 \times 10^{-7}$ |
| Gliese876 c | 30 | 0.13 | 3350 | 0.09 | 0.25 | 0.6 | 0.7 | $2.9 \times 10^{-6}$ |
| HD10647 b | 1003 | 2.03 | 6039 | 0.85 | 2.07 | 0.9 | 0.6 | $9.5 \times 10^{-9}$ |
| HD114729 b | 1135 | 2.08 | 5662 | 1.01 | 2.49 | 0.8 | 0.6 | $9.7 \times 10^{-9}$ |
| HD11506 b | 1270 | 2.43 | 6058 | 1.08 | 2.61 | 3.4 | 0.6 | $6.1 \times 10^{-9}$ |
| HD117618 c | 318 | 0.93 | 5861 | 0.88 | 2.14 | 0.2 | 0.9 | $1.1 \times 10^{-7}$ |
| HD125612 b | 502 | 1.37 | 5897 | 0.78 | 1.91 | 3.0 | 0.6 | $2.2 \times 10^{-8}$ |
| HD12661 b | 264 | 0.83 | 5742 | 0.80 | 1.95 | 2.3 | 0.7 | $7.3 \times 10^{-8}$ |
| HD137388 b | 330 | 0.89 | 5240 | 0.49 | 1.24 | 0.2 | 1.0 | $1.4 \times 10^{-7}$ |
| HD141399 d | 1070 | 2.13 | 5360 | 0.92 | 2.28 | 1.2 | 0.6 | $8.4 \times 10^{-9}$ |
| HD141937 b | 653 | 1.52 | 5925 | 0.80 | 1.94 | 9.7 | 0.6 | $1.6 \times 10^{-8}$ |
| HD142415 b | 386 | 1.05 | 5834 | 0.75 | 1.84 | 1.6 | 0.7 | $4.3 \times 10^{-8}$ |
| HD147513 b | 528 | 1.32 | 5701 | 0.70 | 1.72 | 1.2 | 0.6 | $2.4 \times 10^{-8}$ |
| HD153950 b | 499 | 1.28 | 6076 | 1.05 | 2.55 | 2.7 | 0.7 | $2.9 \times 10^{-8}$ |
| HD154857 b | 409 | 1.29 | 5445 | 1.14 | 2.82 | 2.2 | 0.7 | $2.9 \times 10^{-8}$ |
| HD156411 b | 842 | 1.88 | 5900 | 1.61 | 3.93 | 0.7 | 0.7 | $1.5 \times 10^{-8}$ |
| HD164509 b | 282 | 0.88 | 5922 | 0.79 | 1.94 | 0.5 | 0.7 | $7.2 \times 10^{-8}$ |
| HD169830 c | 2102 | 3.60 | 6266 | 1.52 | 3.66 | 4.0 | 0.6 | $2.7 \times 10^{-9}$ |
| HD196885A b | 1326 | 2.60 | 6340 | 1.51 | 3.62 | 3.0 | 0.6 | $6.2 \times 10^{-9}$ |
| HD202206 b | 256 | 0.83 | 5750 | 0.73 | 1.78 | 17.4 | 0.6 | $6.1 \times 10^{-8}$ |
| HD2039 b | 1183 | 2.20 | 5947 | 0.91 | 2.23 | 4.9 | 0.6 | $7.5 \times 10^{-9}$ |

---

3 It is worth noting that this kind of empirical relation is only indicative as it does not take into account the internal structure of the planet and thus should only be taken as indicative.

| Planet Name | Period [days] | a [AU] | Teff [K] | $r_i$ [AU] | $r_o$ [AU] | Mass [MJup] | Radius [RJup] | Flux Ratio |
|---|---|---|---|---|---|---|---|---|
| HD216435 b | 1311 | 2.56 | 5767 | 1.43 | 3.51 | 1.3 | 0.6 | $6.5\times10^{-9}$ |
| HD218566 b | 226 | 0.69 | 4820 | 0.44 | 1.13 | 0.2 | 1.0 | $2.2\times10^{-7}$ |
| HD23079 b | 626 | 1.50 | 5848 | 0.83 | 2.03 | 2.5 | 0.6 | $1.8\times10^{-8}$ |
| HD28254 b | 1116 | 2.15 | 5664 | 1.03 | 2.53 | 1.2 | 0.6 | $9.6\times10^{-9}$ |
| HD30562 b | 1157 | 2.30 | 5861 | 1.21 | 2.94 | 1.3 | 0.6 | $8.5\times10^{-9}$ |
| HD33564 b | 388 | 1.10 | 6250 | 0.91 | 2.18 | 9.1 | 0.6 | $3.5\times10^{-8}$ |
| HD43197 b | 328 | 0.92 | 5508 | 0.66 | 1.63 | 0.6 | 0.7 | $6.5\times10^{-8}$ |
| HD44219 b | 472 | 1.19 | 5752 | 0.94 | 2.31 | 0.6 | 0.7 | $3.8\times10^{-8}$ |
| HD564 b | 492 | 1.20 | 5902 | 0.75 | 1.84 | 0.3 | 1.2 | $1.2\times10^{-7}$ |
| HD69830 d | 197 | 0.63 | 5385 | 0.57 | 1.41 | 0.1 | 0.5 | $6.4\times10^{-8}$ |
| HD7199 b | 615 | 1.36 | 5386 | 0.61 | 1.51 | 0.3 | 1.2 | $8.2\times10^{-8}$ |
| HD73526 c | 378 | 1.05 | 5590 | 1.01 | 2.49 | 2.5 | 0.7 | $4.5\times10^{-8}$ |
| HD86264 b | 1475 | 2.86 | 6210 | 1.53 | 3.69 | 7.0 | 0.6 | $4.8\times10^{-9}$ |
| HIP57050 b | 41 | 0.16 | 3190 | 0.09 | 0.26 | 0.3 | 1.2 | $5.7\times10^{-6}$ |
| WASP-41 c | 421 | 1.07 | 5546 | 0.67 | 1.67 | 3.2 | 0.6 | $3.6\times10^{-8}$ |
| WASP-47 c | 572 | 1.36 | 5576 | 0.77 | 1.91 | 1.2 | 0.6 | $2.3\times10^{-8}$ |
| muAra b | 643 | 1.50 | 5700 | 0.87 | 2.15 | 1.7 | 0.6 | $1.9\times10^{-8}$ |
| muAra d | 311 | 0.92 | 5700 | 0.87 | 2.15 | 0.5 | 0.7 | $6.5\times10^{-8}$ |
| upsAn d d | 1281 | 2.55 | 6212 | 1.33 | 3.20 | 10.2 | 0.6 | $5.5\times10^{-9}$ |

Table 2: Planets in the exoplanet.eu database within their host star HZ as computed from Equation 2. The last column shows the maximum planet-to-star flux ratio, i.e., at opposition, computed from Equation 1.

From the computed planet-to-star flux ratio we estimated the minimum integration time required for a 3-sigma detection of the orbiting planets. Although planet-to star flux ratios of the order of $10^{-7}$-$10^{-8}$ might seem discouraging, and beyond the capabilities of current generation of observing facilities, this level of precision should be obtainable with the next generation of observatories with 30m class telescopes.

To compute the total exposure time to detect the planets, we used ESO's online Spectroscopic Exposure Time Calculator[4] (ETC) to estimate the SN ($SN_{ref}$) after a 30s exposure observation of a star with a visual magnitude $m_{ref}$ = 6.0 with a fiber fed spectrograph (all settings are presented in Table 3). Please note that the ETC assumes a theoretical efficiency value of 25% for the telescope+instrument+detector. Different instuments attached to the E-ELT will affect the total efficiency and thus these values should only be taken as indicative. A more detailed description of the ETC can be found on the online documentation[5].

| Parameter | Value |
|---|---|
| **Target Input Flux Distribution** | |
| *Spectral Type* | G2V |
| *Vega mag* | V = 6.0 |
| *Spatial distribution* | Point source |
| **Telescope Setup** | |
| *Observatory site* | Paranal (2635 m) |
| *Telescope diameter* | 39 m |
| **Sky Conditions** | |
| *Seeing* | 0.8 arcsec |
| *Airmass* | 1.15 |
| **Instrument Setup** | |
| *AO mode* | Seeing limited |
| *Observing wavelength* | Band: V (550nm) |

---

4 https://www.eso.org/observing/etc/bin/gen/form?INS.NAME=E-ELT+INS.MODE=swspectr
5 https://www.eso.org/observing/etc/doc/elt/etc_spec_model.pdf

| | |
|---:|:---|
| *Radius of circular S/N area* | 1000.0 mas |
| *Number of spectra on the detector* | 1 |
| *Spectral resolution* | 100000 |
| **Results** | |
| *Exposure time* | NDIT = 1 ; DIT = 30s |

Table 3: Settings used in the EELT ETC.

Then, for each target in our sample, we used Equation 5 to extrapolate the signal-to-noise ($SN_{host}$) of a single 30s exposure of the target's host from its visual magnitude ($m_{host}$).

$$m_{host} - m_{ref} = -5 \log \left( \frac{SN_{host}}{SN_{ref}} \right) \quad (5)$$

The required total exposure time required for a 3-sigma detection of the target will be given by Equation 6.

$$t_{3\sigma} = \frac{t_{ref}}{N_{lines}} \left( \frac{SN_{planet}}{SN_{host}} \right)^2 \quad (6)$$

The results for all planets in the sample are shown in Figure 4. Table 4 shows the cases where a 3-sigma detection can be accomplished under 100h (in green on Figure 4)..

| Planet name | Spectral Type of host star | $SN_{Planet}$ 3-sigma | $Mag_V$ | SN after 30s | Required Total Exposure Time |
|---|---|---|---|---|---|
| 55Cnc f | K0IV-V | 2.68e+07 | 6.0 | 4031.8 | 89.5 |
| Gl687 b | M3.5V | 3.05e+06 | 9.2 | 923.6 | 22.1 |
| Gliese876 b | M4V | 3.54e+06 | 10.2 | 577.4 | 76.0 |
| Gliese876 c | M4V | 1.05e+06 | 10.2 | 577.4 | 6.7 |
| HIP57050 b | M4V | 5.28e+05 | 11.9 | 260.3 | 8.3 |

Table 4: Required exposure time for the planets in Table 2 that require less than 100h for a 3-sigma detection.

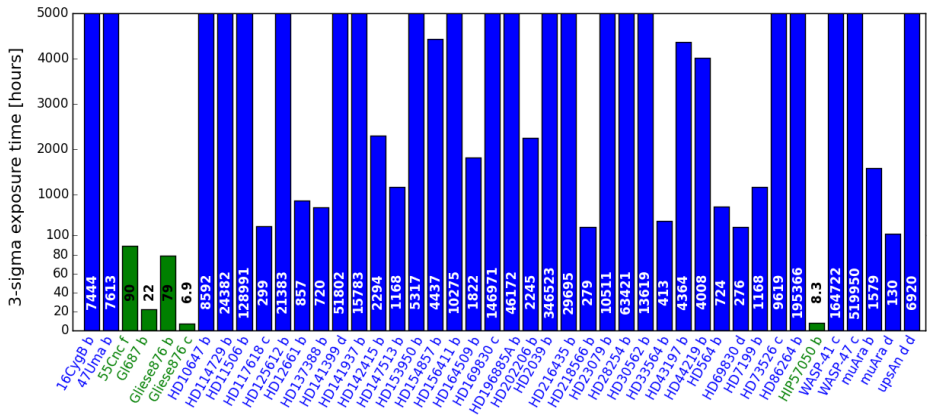

Figure 4: The results for all planets in the sample.

Please note that this analysis we assume that the FWHM of the reflected planetary CCF will be the same as the FWHM of its host star. For example, a high rotation rate of the planet will cause the reflected to be broadened relatively to the one of the star (this effect has been show to be useful towards the detection of the rotation velocity of the planet, see Kawahara 2012) while reducing its amplitude, making it harder to detect. An additional effect that might change the FWHM of the detected signal is that the planet will reflect the stellar signature as seen from the planet rest frame (see Collier Cameron et al. 2002 for more details). For example, a planet whose orbital period is the same as its host rotation period will reflect the host's integrated spectral signature unbroadened due to stellar rotation as it always "sees" the same face of the star. Meanwhile an observer on Earth will observe the host's spectral signature broadened due to rotation. As a practical example, on Earth we see a solar v sin i of ~1.6 km/s instead of ~1.8 km/s that we would observe if we were not orbiting the Sun.

# 4 Discussion and conclusions

As can be seen from Equation 1, the main problem faced by researchers when attempting to detect and characterize exoplanets in the optical is the low planet-to-star flux ratio. Even in the best scenarios, e.g. giant planets in close orbits, this flux ratio does not go above $10^{-4}$. As the planet orbital period increases, this ratio decreases further. To test for the detection of planets in their host's habitable zones, the limits of the habitable zone of all host stars in the exoplanet.eu database (Schneider et al. 2011) were computed using the model of Kasting, Whitmire & Reynolds 1993. We limited the initial sample to planets whose main sequence stars were brighter than $mag_V=10$ and had effective temperatures in the 2000K-8000K range. The test sample was selected as the planets in the initial sample whose semi-major axis lies within the limits of their host's habitable zones. For each of the planets in the sample the planet-to-star flux ratio was computed using Equation 1 (See Table 1). Even in the most favourable case, this flux ratio does not go above $2 \times 10^{-7}$, yielding a SN for a 3-sigma detection equals to $1.5 \times 10^7$, a SN level not available to current observing facilities.

To surpass this problem, we proposed to use a technique that makes use of the Cross Correlation Function to increase the SN of spectrographic observations obtained with next generation Extremely Large Telescopes (e.g. ESO's E-ELT). This technique has already been tested successfully to recover the reflected signature of the prototypical hot-Jupiter 51 Pegasi b from HARPS observations (see Martins et al. 2015). Our proposal is that this technique, combined with observation using a high-resolution spectrograph fed by the huge collecting power of the E-ELT should allow to achieve the SN level required to detect the planets in the habitable zone of their host star. To do so, the total exposure time necessary to achieve the required SN level for each planet in the sample was computed (see Figure 4). Of those, 5 targets have total exposure times under 100h, making them good test candidates for the detection of their spectrographic reflected signals. Although the final sample we present is composed only of giant planets, those might be orbited by satellites with conditions to have liquid water on their surface and support life as we know it, yielding good prospects towards the detection of a habitable planet similar to our own.

It is worth noting that all shortlisted FJK targets have magnitudes brighter than 6, indicating that even with the EELT only habitable planets orbiting bright solar-type stars will be detectable. Interesting is the case of the remaining targets, planets orbiting M dwarfs: although the hosts have low brightness, the habitable zone is close enough to the star to make these planets detectable with the EELT. Of particular interest are the cases of Gliese876 c and is HIP57050 b as they are close enough to their host to be detected with around or less than 8h of total integration time. If we extrapolate to a VLT sized telescope, this will mean a 3-sigma detection under 40h, meaning that even current generation facilities can start to be used to detect these targets.

Our simple approach shows that combining the power of the Cross Correlation Function with observations from a high-resolution spectrograph fed by the ESO's E-ELT, the detection of the reflected spectral signature of exoplanets in their host's habitable zones is indeed possible. Of great importance towards these studies are surveys missions like PLATO ($4 < mag_V < 16$) who will be able to provide a large amount of suitable targets to this genre of study.

**Aknowlegments:** *This work was supported by the COST Action TD 1308 «ORIGINS» and by Fundação para a Ciência e a Tecnologia (FCT) through research grants UID/FIS/04434/2013 (POCI-01-0145-FEDER-007672) and project PTDC/FIS-AST/1526/2014. PF and NCS acknowledge support by Fundação para a Ciência e a Tecnologia (FCT) through Investigador FCT contracts of reference IF/01037/2013 and IF/00169/2012, respectively, and POPH/FSE (EC) by FEDER funding through the program "Programa Operacional de Factores de Competitividade - COMPETE". PF further acknowledges support from Fundação para a Ciência e a Tecnologia (FCT) in the form of an exploratory project of reference IF/01037/2013CP1191/CT0001.*